\documentclass[final,5p,times]{elsarticle}

\biboptions{sort&compress}

\usepackage{times}
\usepackage{euscript}
\usepackage[fleqn]{amsmath}
\usepackage{amssymb}
\usepackage{amsfonts}
\usepackage{graphicx}
\usepackage{color}

\journal{Nuclear Instruments and Methods in Phys. Res. B}

\begin{document}

\begin{frontmatter}

\title{Dynamic screening of an ion in a degenerate electron gas\\
within the second-order Born approximation}

\author[label1,label2]{Hrachya~B.~Nersisyan\corref{cor1}}
\ead{hrachya@irphe.am}

\author[label3]{Jos\'{e}~M.~Fern\'{a}ndez-Varea}

\author[label4]{N\'{e}stor~R.~Arista}

\address[label1]{Plasma Theory Group,
Institute of Radiophysics and Electronics,
0203 Ashatarak, Armenia}

\address[label2]{Center of Strong Fields Physics,
Yerevan State University,
Alex Manoogian str.\ 1, 0025 Yerevan, Armenia}

\address[label3]{Facultat de F\'{\i}sica (ECM and ICC),
Universitat de Barcelona,
Diagonal~645, E-08028 Barcelona, Spain}

\address[label4]{Divisi\'{o}n de Colisiones At\'{o}micas,
Centro At\'{o}mico Bariloche and Instituto Balseiro,
Comisi\'{o}n Nacional de Energ\'{\i}a At\'{o}mica,
8400 Bariloche, Argentina}

\cortext[cor1]{Corresponding author:
Tel.: +374~10~287850; fax: +374~232~33770}

\begin{abstract}

The dynamic Friedel sum rule (FSR) is derived within the second-order
Born (B2) approximation for an ion that moves in a fully degenerate
electron gas and for an arbitrary spherically-symmetric electron-ion
interaction potential. This results in an implicit equation for the
dynamic B2 screening parameter which depends on the ion atomic number
$Z_{1}$ unlike the first-order Born (B1) dynamic screening parameter
reported earlier by some authors. Furthermore, for typical metallic
densities our analytical results for the Yukawa and hydrogenic
potentials are compared, for both positive and negative ions, to the
exact screening parameters calculated self-consistently by imposing the
exact dynamic FSR requirement to the scattering phase shifts. The B1 and
B2 screening parameters agree excellently with the exact values at large
velocities, while at moderate and low velocities the B1 approximation
deviates from the exact solution whereas the B2 approximation still
remains close to it. In addition, a Pad\'{e} approximant to the Born
series yields a further improvement of the perturbative approach,
showing an excellent agreement on the whole velocity range in the case
of antiprotons.

\end{abstract}


\end{frontmatter}


\section{Introduction}
\label{sec:int}

The dynamic screening of swift heavy charged particles in condensed
matter is a phenomenon that is important to understand the electronic
stopping power and related projectile-target interaction properties. The
screening experienced by an intruder charge arises from the electron
density induced in the traversed medium and affects the stopping
properties of the particles. It is therefore of interest to determine
how this screening effect varies with the projectile velocity.

An approach commonly used to describe dynamic
screening effects is based on the dynamic Friedel sum rule (FSR)
proposed in \cite{nag96,lif98} which uses the concept of the shifted
Fermi sphere \cite{zar95}. This rule is very useful to adjust in a
self-consistent way the electron-ion interaction potential and the
related screening length. In this paper we study the dynamic FSR for a
point-like ion that moves in a fully degenerate electron gas (DEG)
within the framework of the second-order Born (B2) approximation. The
Born approximation has been previously used in conjunction with the
static \cite{ech89} or dynamic FSR \cite{nag96,lif98}, but only within
the first-order Born (B1) approximation. This is somewhat unsatisfactory
because the resulting B1 screening length is independent of the ion
atomic number and is therefore identical for a particle and its
antiparticle. Recently, the static screening length has been deduced
within the B2 approximation \cite{ner05,ner13}. The static B2 screening
lengths pertaining to protons and antiprotons agree satisfactorily with
the exact numerical solutions at electron densities typical of metals.
In this context, our main purpose is to go beyond the B1 approximation,
and consider the B2 approximation for the dynamic FSR. This generalizes
the previous results obtained within the static B1 \cite{ech89} or B2
\cite{ner05,ner13} and the dynamic B1 approximations \cite{nag96,lif98}
and hence furnishes useful numerical estimates of the influence of both
the ion charge and its velocity on the screening length in a DEG.


\section{Self-consistent formulation of the dynamic FSR}
\label{sec:2}

Let us revisit the dynamic FSR first formulated by Nagy and Bergara
\cite{nag96} and later studied in more detail by Lifschitz and Arista
\cite{lif98} for a DEG. To this end, consider an ion with charge
$Z_{1}e$ ($Z_{1}$ is the ion atomic number)
and constant velocity $\mathbf{v}$ that moves through a DEG of
density $n_{\mathrm{e}}$ [Fermi wave number
$k_{\mathrm{F}}=(3\pi^{2}n_{\mathrm{e}})^{1/3}$]. An electron whose wave
vector is $\mathbf{k}_{\mathrm{e}}$ collides elastically with the ion.
The relative velocity of the colliding particles is denoted as
$\mathbf{v}_{\mathrm{r}}=\mathbf{v}_{\mathrm{e}}-\mathbf{v}$, where
$\mathbf{v}_{\mathrm{e}}=\hbar\mathbf{k}_{\mathrm{e}}/m_{\mathrm{e}}$
is the electron's initial velocity.
The relative wave vector is
$\mathbf{k}_{\mathrm{r}}=\mathbf{k}_{\mathrm{e}}-\mathbf{k}$ with
$\mathbf{k}=m_{\mathrm{e}}\mathbf{v}/\hbar$ (note that $\mathbf{k}$ is
not the wave vector of the ion). In the center of mass (c.m.) frame of
reference the wave function of the incoming free electron is
$\phi_{\mathbf{k}_{\mathrm{r}}}(\mathbf{r})=
\mathrm{e}^{\mathrm{i}\mathbf{k}_{\mathrm{r}}\cdot\mathbf{r}}$ whereas
its wave function after the collision is given by the partial-wave
expansion (see, e.g., \cite{lan81})
\begin{equation}
\psi_{\mathbf{k}_{\mathrm{r}}}(\mathbf{r}) =
\sum_{\ell=0}^{\infty} \mathrm{i}^{\ell} \, (2\ell+1) \,
\mathrm{e}^{\mathrm{i}\delta_{\ell}(k_{\mathrm{r}})} \,
\Re_{k_{\mathrm{r}},\ell}(r) \, P_{\ell}(\cos\theta),
\label{eq:1}
\end{equation}
where $\Re_{k_{\mathrm{r}},\ell}(r)$ and $\delta_{\ell}(k_{\mathrm{r}})$
are, respectively, the radial wave function and the scattering phase
shifts corresponding to the angular momentum $\ell$ and depending only
on $k_{\mathrm{r}}$ (the modulus of $\mathbf{k}_{\mathrm{r}}$), $\theta$
is the scattering angle in the c.m.\ reference frame (i.e., the angle
between $\mathbf{k}_{\mathrm{r}}$ and $\mathbf{r}$), and $P_{\ell}$ are
the Legendre polynomials.

Following \cite{nag96} we introduce now the electron density induced in
the DEG by the moving ion
\begin{equation}
n_{\mathrm{ind}}(\mathbf{r}) =
\frac{2}{(2\pi)^{3}} \int_{k_{\mathrm{e}}\leqslant k_{\mathrm{F}}}
\left(
|\psi_{\mathbf{k}_{\mathrm{r}}}(\mathbf{r})|^{2}
- |\phi_{\mathbf{k}_{\mathrm{r}}}(\mathbf{r})|^{2}
\right)
\mathrm{d}\mathbf{k}_{\mathrm{e}},
\label{eq:2}
\end{equation}
where the integration is performed in the domain
$k_{\mathrm{e}}\leqslant k_{\mathrm{F}}$ as a consequence of the Pauli
exclusion principle. Let us stress that $n_{\mathrm{ind}}(\mathbf{r})$
is not isotropic because of the motion of the ion. This fact is
expressed in averaging of the induced density with respect to the
unit-step distribution function of a DEG in the laboratory frame of
reference (wave vector $\mathbf{k}_{\mathrm{e}}$) while
$|\psi_{\mathbf{k}_{\mathrm{r}}}(\mathbf{r})|^{2}$ and
$|\phi_{\mathbf{k}_{\mathrm{r}}}(\mathbf{r})|^{2}$ are calculated in the
c.m.\ reference frame. In the case of an ion at rest
$\mathbf{k}_{\mathrm{r}}=\mathbf{k}_{\mathrm{e}}$ and Eq.~(\ref{eq:2})
becomes the equation addressed by Friedel in \cite{fri52}, which yields
the well-known static FSR \cite{fri52} and an isotropic induced electron
density.

Next we calculate the total charge induced in a spherical volume
$\Omega_{R}$ around the ion, which is
$Q_{\mathrm{ind}}=-eN_{\mathrm{ind}}(v)$ with
\begin{equation}
N_{\mathrm{ind}}(v) =
\int_{\Omega_{R}} n_{\mathrm{ind}}(\mathbf{r}) \, \mathrm{d}\mathbf{r} =
\frac{2}{(2\pi)^{2}} \int_{k_{\mathrm{e}}\leqslant k_{\mathrm{F}}}
A(k_{\mathrm{r}}) \,
\frac{\mathrm{d}\mathbf{k}_{\mathrm{e}}}{k_{\mathrm{r}}^{2}},
\label{eq:3}
\end{equation}
where $R$ (with $R\to\infty$) is the radius of the volume $\Omega_{R}$ and
\begin{equation}
A(k_{\mathrm{r}}) =
\frac{k_{\mathrm{r}}^{2}}{2\pi} \int_{\Omega_{R}}
\left(
|\psi_{\mathbf{k}_{\mathrm{r}}}(\mathbf{r})|^{2}
- |\phi_{\mathbf{k}_{\mathrm{r}}}(\mathbf{r})|^{2}
\right)
\mathrm{d}\mathbf{r};
\label{eq:4}
\end{equation}
notice that the function $A(k_{\mathrm{r}})$ differs from the definition
adopted in \cite{ser71,nag96} by a factor $k_{\mathrm{r}}^{2}/2\pi$.
From Eqs.~(\ref{eq:1}) and (\ref{eq:4}) it is seen that
$A(k_{\mathrm{r}})$ is isotropic and depends only on $k_{\mathrm{r}}$.
This enables the angular integration in Eq.~(\ref{eq:3}) which results
in
\begin{eqnarray}
N_{\mathrm{ind}}(v)
& = &
\frac{2}{\pi}
\left\{
\Theta(k_{\mathrm{F}}-k) \int_{0}^{k_{\mathrm{F}}-k} A(q) \, \mathrm{d}q
\right.
\nonumber
\\
& &
\left.
\mbox{}
+ \frac{1}{4k} \int_{|k-k_{\mathrm{F}}|}^{k+k_{\mathrm{F}}}
\left[ k^{2}_{\mathrm{F}} - (k-q)^{2} \right] A(q) \, \frac{\mathrm{d}q}{q}
\right\},
\label{eq:5}
\end{eqnarray}
$\Theta(\kappa)$ is the Heaviside unit-step function. The quantity
$A(q)$ can be evaluated in closed form using Servadio's general
relation \cite{ser71}. When $R\to\infty$ this function can be expressed
through the scattering amplitude $f(q,\theta)$ as follows
\begin{eqnarray}
A(q)
& = &
\frac{\partial}{\partial q}\left[qf^{\ast}(q,0)\right]
\nonumber
\\
& &
\mbox{}
+ \mathrm{i} \int_{0}^{\pi} qf(q,\theta) \,
\frac{\partial}{\partial q} \left[ qf^{\ast}(q,\theta) \right]
\sin\theta \, \mathrm{d}\theta
\label{eq:6}
\\
& = &
\sum_{\ell=0}^{\infty} (2\ell+1) \, \delta_{\ell}'(q).
\label{eq:7}
\end{eqnarray}
Here $f(q,0)$ is the scattering amplitude for $\theta=0$, the asterisc
denotes complex conjugation and the prime indicates derivation with
respect to the argument. The second part of Servadio's relation,
Eq.~(\ref{eq:7}), is easily found by substitution of the partial-wave
expansion of the scattering amplitude (see, e.g., \cite{lan81}) into
Eq.~(\ref{eq:6}).

At this point we impose the condition that the intruder
ion has to be completely screened at large distances,
$Z_{1}e+Q_{\mathrm{ind}}=0$, which serves as the basic constraint for
the scattering theory. It was first suggested by Friedel \cite{fri52}
and can be viewed as the conservation of the total charge of a
many-electron system. In this sense the FSR is similar to the optical
theorem of scattering theory \cite{lan81} which requires the
conservation of particle number (for inelastic scattering the number of
the particles participating in the elastic scattering). Using
Eq.~(\ref{eq:5}) the condition of complete screening can be rewritten in
the explicit form
\begin{equation}
Z_{1} = N_{\mathrm{ind}}(v)
\label{eq:8}
\end{equation}
with $v=\hbar k/m_{\mathrm{e}}$. If the projectile carries
with it $N_{\mathrm{b}}$ bound electrons, in Eq.~(\ref{eq:8})
one should simply replace $Z_{1}$ with $Z_{1}-N_{\mathrm{b}}$ \cite{fri52}.
In order to represent the dynamic FSR,
Eq.~(\ref{eq:8}), in a more familiar form (as a sum over partial waves)
we insert Eq.~(\ref{eq:7}) into Eq.~(\ref{eq:5}) and integrate by parts
assuming that $\delta_{\ell}(0)=0$, which yields
\begin{equation}
Z_{1} =
\frac{2}{\pi} \sum_{\ell=0}^{\infty} (2\ell+1) \, \Delta_{\ell}(v)
\label{eq:9}
\end{equation}
with the ``dynamic phase shifts''
\begin{equation}
\Delta_{\ell}(v) =
\frac{1}{4k} \int_{\vert k-k_{\mathrm{F}}\vert}^{k+k_{\mathrm{F}}}
\left( 1 + \frac{k_{\mathrm{F}}^{2}-k^{2}}{q^{2}} \right)
\delta_{\ell}(q) \, \mathrm{d}q.
\label{eq:10}
\end{equation}
Eqs.~(\ref{eq:9}) and (\ref{eq:10}) are identical to the ``extended''
FSR of Lifschitz and Arista~\cite{lif98} for a DEG, which was deduced
having recourse to geometrical arguments about the Galilean
transformation of the Fermi sphere. The extension of the dynamic FSR to
an electron gas at high temperature has been outlined by Nagy and
Bergara \cite{nag96}.

It is straightforward to see that in the limit $v\to 0$
Eq.~(\ref{eq:10}) reduces to
$\Delta_{\ell}(0)=\delta_{\ell}(k_{\mathrm{F}})$ which together with
Eq.~(\ref{eq:9}) constitutes the static FSR \cite{fri52}. In the
high-velocity limit, within the leading order from Eq.~(\ref{eq:10}) one
gets
$\Delta_{\ell}(v)=(v_{\mathrm{F}}^{2}/3v^{2})k_{\mathrm{F}}\delta_{\ell}'(k)$,
where $v_{\mathrm{F}}=\hbar k_{\mathrm{F}}/m_{\mathrm{e}}$ is the Fermi
velocity \cite{lif98}. Interestingly, the dynamic phase shifts are
expressed in the high-velocity regime through the momentum derivative of
the ordinary phase shifts.


\section{First- and second-order Born approximations}
\label{sec:3}

With the theoretical formalism presented in Section~\ref{sec:2}, we take
up the main topic of this paper, namely to study the dynamic FSR for a
point-like ion moving in a DEG within up to the B2 approximation. Hence
we look for the scattering amplitude in Eq.~(\ref{eq:6}) in a
perturbative manner writing $f=f_{\mathrm{B1}}+f_{\mathrm{B2}}$, where
$f_{\mathrm{B1}}$ and $f_{\mathrm{B2}}$ are the first- and second-order
scattering amplitudes, respectively. Similarly, we expand Servadio's
function perturbatively to the second order,
$A=A_{\mathrm{B}1}+A_{\mathrm{B}2}$. Introducing in Eq.~(\ref{eq:7}) the
corresponding expansion of the phase shifts,
$\delta_{\ell}=\delta_{\ell,\mathrm{B}1}+\delta_{\ell,\mathrm{B}2}$, we
get with the help of Eq.~(19) in \cite{ner13}
\begin{eqnarray}
A_{\mathrm{B}1}(q)
& = &
\frac{\partial}{\partial q} \big\{ q f_{\mathrm{B1}}(q,0) \big\},
\label{eq:11}
\\
A_{\mathrm{B}2}(q)
& = &
\frac{\partial}{\partial q} \big\{ q \, \mathrm{Re}[f_{\mathrm{B2}}(q,0)] \big\},
\label{eq:12}
\end{eqnarray}
where $f_{\mathrm{B1}}(q,0)$ and $f_{\mathrm{B2}}(q,0)$ are the B1 and
B2 forward-scattering amplitudes. Then, using Eqs.~(23) and (26) in
\cite{ner13} we arrive at
\begin{eqnarray}
A_{\mathrm{B1}}(q)
& = &
-\frac{m_{\mathrm{e}}}{2\pi\hbar^{2}} \, \widetilde{V}(0),
\label{eq:13}
\\
A_{\mathrm{B2}}(q)
& = &
\frac{4m^{2}_{\mathrm{e}}}{(2\pi)^{3}\hbar^{4}}
\int_{0}^{\infty}
\widetilde{V}^{2}(\kappa) \,
\frac{\kappa^{2}\,\mathrm{d}\kappa}{\kappa^{2}-4q^{2}},
\label{eq:14}
\end{eqnarray}
where $\widetilde{V}(q)$ is the Fourier transform of the electron-ion
interaction potential $V(r)$ given by
\begin{equation}
\widetilde{V}(q) =
\int_{0}^{\infty} V(r) \, j_{0}(qr) \, 4\pi r^{2} \, \mathrm{d}r
\label{eq:15}
\end{equation}
with $j_{0}(z)=\sin z/z$.

Next we formulate the dynamic FSR within the B2 approximation. In this
approximation $Z_{1}=N_{\mathrm{ind,B1}}(v)+N_{\mathrm{ind,B2}}(v)$,
where $N_{\mathrm{ind,B1}}(v)$ and $N_{\mathrm{ind,B2}}(v)$ are the
first- and second-order quantities corresponding to the function
$N_{\mathrm{ind}}(v)$ in Eq.~(\ref{eq:5}) and involving
$A_{\mathrm{B1}}(q)$ and $A_{\mathrm{B2}}(q)$, respectively. The ensuing
equation is the main result of the present article. The first and second
terms in this relation are linear ($\sim Z_{1}$) and quadratic ($\sim
Z_{1}^{2}$) with respect to the interaction potential and represent,
respectively, the first- and second-order Born contributions to the
dynamic FSR. While the first-order term has been derived previously
\cite{nag96} the second one is a new result that may be regarded as the
counterpart to the Barkas--Andersen correction in the Bethe--Bloch
stopping power formula. In general, the obtained equation is a
transcendental equation that serves to determine the dynamic screening
length $\lambda(v)\equiv 1/\alpha(v)$, where $\alpha(v)$ is the
corresponding dynamic screening parameter, of the interaction potential
involved in $A_{\mathrm{B1}}(q)$ and $A_{\mathrm{B2}}(q)$.

In what follows we apply the dynamic FSR in the B2 approximation to the
very important group of screened potentials
\begin{equation}
V(r) =
-\frac{Z_{1}e^{2}}{r} \, \Phi(\alpha r),
\label{eq:16}
\end{equation}
where $\Phi(x)$ is the screening function. Then
\begin{equation}
\widetilde{V}(q) =
-\frac{4\pi Z_{1}e^{2}}{\alpha^{2}} \,
\widetilde{\mathcal{V}}(q/\alpha),
\label{eq:17}
\end{equation}
where
\begin{equation}
\widetilde{\mathcal{V}}(x) =
\int_{0}^{\infty} \Phi(y) j_{0}(xy) \, y \, \mathrm{d}y
\label{eq:18}
\end{equation}
is the dimensionless Fourier transform of the interaction potential. In
terms of $\widetilde{\mathcal{V}}(x)$ we have
$\widetilde{V}(0)=-(4\pi Z_{1}e^{2}/\alpha^{2})\gamma$ with (see
\cite{ner13})
\begin{equation}
\gamma =
\widetilde{\mathcal{V}}(0) =
\int_{0}^{\infty} \Phi(x) \, x \, \mathrm{d}x.
\label{eq:19}
\end{equation}
Now we substitute Eqs.~(\ref{eq:17}) and (\ref{eq:18}) for an
arbitrary screened potential into Eqs.~(\ref{eq:13}) and (\ref{eq:14}),
and these into the B2 dynamic FSR, then integrating over $q$. It turns
out that the general solution for the screening parameter $\alpha$ in
the B2 approximation can be cast in the implicit form
\begin{equation}
\alpha(v) =
\alpha_{\mathrm{RPA}}
\left[ \gamma \, F(s) + \frac{\pi\chi^{2}}{2} \, Z_{1} \, G(s,u) \right]^{1/2}.
\label{eq:20}
\end{equation}
Here
$\alpha_{\mathrm{RPA}}=(4k_{\mathrm{F}}/\pi a_{0})^{1/2}=1/\lambda_{\mathrm{TF}}$,
where $\lambda_{\mathrm{TF}}$ is the Thomas--Fermi screening length,
$u=2k_{\mathrm{F}}/\alpha$, $\chi^{2}=(\pi k_{\mathrm{F}}a_{0})^{-1}$ is
the (dimensionless) Lindhard density parameter of the DEG, $a_{0}$ is
the Bohr radius. Besides, the functions $F(s)$ and $G(s,u)$ in
Eq.~(\ref{eq:20}) depend on the ion velocity through
$s=v/v_{\mathrm{F}}$ and are given by
\begin{equation}
F(s) =
\frac{1}{2} + \frac{1-s^{2}}{4s} \, \ln \left| \frac{1+s}{1-s} \right|,
\label{eq:21}
\end{equation}
and
\begin{equation}
G(s,u) =
\frac{u}{\pi} \int_{0}^{\infty} Q(s,x/u) \,
\widetilde{\mathcal{V}}^{2}(x) \, \mathrm{d}x
\label{eq:22}
\end{equation}
with
\begin{eqnarray}
Q(s,a)
& = &
\frac{1-s^{2}}{s} \ln \left| \frac{1+s}{1-s} \right|
+ a \ln \left| \frac{s^{2}-(a+1)^{2}}{s^{2}-(a-1)^{2}} \right|
\nonumber
\\
& &
\mbox{}
+ \frac{s^{2}-1+a^{2}}{2s}
\ln \left\vert \frac{(s+1)^{2}-a^{2}}{(s-1)^{2}-a^{2}} \right\vert.
\label{eq:23}
\end{eqnarray}
The numerical constant $\gamma$ and the function $G(s,u)$ should be
specified for the adopted screened potential according to
Eqs.~(\ref{eq:19}) and (\ref{eq:22}), (\ref{eq:23}), respectively.

In Eq.~(\ref{eq:20}) the term containing the function $G(s,u)$ is the B2
correction to the screening parameter $\alpha$. Neglecting this term we
retrieve the familiar expression
\begin{equation}
\alpha_{\mathrm{B1}}(v) =
\alpha_{\mathrm{RPA}} \, \big[ \gamma F(s) \big]^{1/2}
\label{eq:24}
\end{equation}
for $\alpha$ in the B1 approximation \cite{nag96,lif98}. In contrast to
$\alpha_{\mathrm{B1}}$, the second-order screening parameter
(\ref{eq:20}) depends on $Z_{1}$ and thus predicts different screening
parameters for attractive ($Z_{1}>0$) and repulsive ($Z_{1}<0$)
electron-ion potentials. It should also be noted that $F(s)$ coincides
with the screening function of the Lindhard dielectric function in the
static limit, $\varepsilon_{\mathrm{L}}(q,\omega\to
0)=1+(q\lambda_{\mathrm{TF}})^{-2}F(q/2k_{\mathrm{F}})$ \cite{lin54}.

Equation (\ref{eq:20}) determines the second-order screening parameter
as a function of the density of the DEG and the atomic number and the
velocity of the moving ion. Recalling that the second-order correction
[i.e., the second term in Eq.~(\ref{eq:20})] should be smaller than the
first one, Eq.~(\ref{eq:20}) can be solved iteratively. A simple
estimate is achieved if $\alpha_{\mathrm{B}1}$, Eq.~(\ref{eq:24}), is
inserted into the second-order term of Eq.~(\ref{eq:20}). For typical
densities of conduction electrons in metals ($1.5\lesssim
r_{\mathrm{s}}\lesssim 5$\footnote{The one-electron radius
$r_{\mathrm{s}}$ is proportional to the Lindhard density parameter,
$r_{\mathrm{s}}=(9\pi^{4}/4)^{1/3}\chi^{2}$.}) and for (bare) protons
and antiprotons ($Z_{1}=\pm 1$) the second-order correction is indeed
small.

It is of particular interest to study Eq.~(\ref{eq:20}) at low and high
velocities of the ion. At low velocities ($s\ll 1$), from
Eqs.~(\ref{eq:21}) and (\ref{eq:23}) it is easy to see that $F(s)\simeq 1$
and $Q(s,a)\simeq 2a\ln|(a+1)/(a-1)|$. Introducing the latter
relation in Eq.~(\ref{eq:22}) we recover the function
$G(0,u)\equiv g(u)$ of \cite{ner13}. Thus, in the low-velocity regime
Eq.~(\ref{eq:20}) with $F(0)=1$ and $G(0,u)=g(u)$ coincides with our
previous results deduced within the B2 approximation for a static ion
\cite{ner13}. On the other hand, at high velocities ($s\gg 1$), from
Eqs.~(\ref{eq:21}) and (\ref{eq:23}) it follows that $F(s)\simeq
1/3s^{2}+1/15s^{4}$ and $Q(s,a)\simeq -4a^{2}/3s^{4}$. Inserting the
latter expression in Eq.~(\ref{eq:22}) yields $G(s,u)\simeq
-C_{\infty}/3us^{4}$ with (see \cite{ner13})
\begin{equation}
C_{\infty} =
\frac{4}{\pi}
\int_{0}^{\infty} \widetilde{\mathcal{V}}^{2}(x) \, x^{2} \, \mathrm{d}x =
2 \int_{0}^{\infty} \Phi^{2}(x) \, \mathrm{d}x.
\label{eq:25}
\end{equation}
With the asymptotic formulas for $F(s)$ and $G(s,u)$ it is easy to
evaluate the screening length at large velocities, $v\gg
v_{\mathrm{F}}$, taking into account the B2 corrections. From
Eq.~(\ref{eq:20}) we get
\begin{equation}
\lambda(v) =
\gamma^{-1/2} \frac{v}{\omega_{\mathrm{p}}}
\left[
1 - \frac{v_{\mathrm{F}}^{2}}{10v^{2}}
\left(
1 - \frac{5C_{\infty}Z_{1}e^{2}\omega_{\mathrm{p}}}{8\gamma^{1/2}E_{\mathrm{F}}v}
\right)
\right],
\label{eq:26}
\end{equation}
where $E_{\mathrm{F}}=\frac{1}{2}m_{\mathrm{e}}v_{\mathrm{F}}^{2}$ and
$\omega_{\mathrm{p}}=(4\pi e^{2}n_{\mathrm{e}}/m_{\mathrm{e}})^{1/2}$
are the Fermi energy and the plasma frequency of the DEG, respectively.
The dominant term (taking $\gamma=1$), $v/\omega_{\mathrm{p}}$,
corresponds to the usual behaviour of the dynamic screening of swift
ions in a DEG \cite{lin54}. This contribution shows that collective-like
effects characterize the dynamic screening in the DEG albeit these
effects have not been explicitly included in our treatment; they arise
as a consequence of the imposed self-consistent FSR requirement. A
similar appearance of the collective behaviour in a velocity-dependent
density-functional description has been discussed in \cite{zar95}. The
$\mathcal{O}(v^{-1})$ and $\mathcal{O}(v^{-2})$ terms are the
higher-order velocity corrections to the screening length. In
particular, the last term containing the ion charge $Z_{1}$ is the B2
correction to $\lambda(v)$. As expected, at high velocities ($v\gg
v_{\mathrm{F}}$), the contribution of this term is smaller compared to
the other terms. It is noteworthy that in general the main contribution
in Eq.~(\ref{eq:26}) involves the parameter $\gamma$ which varies
significantly for various interaction potentials (see, e.g.,
\cite{ner13} and Sections~\ref{sec:yuk} and \ref{sec:hyd} below) and
fixes their spatial ranges. Therefore, the Friedel adjustment will give
larger (smaller) screening lengths to compensate for the smaller
(larger) spatial ranges of the potentials.

For practical applications we provide explicit expressions for the
Yukawa and hydrogenic potentials, which are often employed to model the
stopping of ions with either quantum or classical formalisms
\cite{fer77,apa87,ven88,sor90,cal94,nag99,ari02,ari04,ari06}. The
numerical constants $\gamma$ and $C_{\infty}$ for these interaction
potentials have been evaluated in \cite{ner13}, and in
Sections~\ref{sec:yuk} and \ref{sec:hyd} we present the respective
functions $G(s,u)$.

\subsection{Yukawa potential}
\label{sec:yuk}

The screening function of the Yukawa potential is
$\Phi(x)=\mathrm{e}^{-x}$ and its Fourier transform reads
\begin{equation}
\widetilde{V}(q) =
-\frac{4\pi Z_{1}e^{2}}{q^{2}+\alpha^{2}}.
\label{eq:27}
\end{equation}
Substituting $\Phi(x)$ and $\widetilde{V}(q)$ into
Eqs.~(\ref{eq:19}), (\ref{eq:22}) and (\ref{eq:25}) we arrive
at $\gamma=1$ \cite{ech89,lif98,ner13},
\begin{equation}
G(s,u) =
\frac{1}{8su}
\left[
(\eta\xi+1) \ln \frac{\eta^{2}+1}{\xi^{2}+1}
- 2\eta\xi \ln \frac{\eta}{|\xi|}
\right]
\label{eq:28}
\end{equation}
and $C_{\infty}=1$ \cite{ner13}, respectively,
where $\xi=u(s-1)$ and $\eta=u(s+1)$.

\subsection{Hydrogenic potential}
\label{sec:hyd}

For the hydrogenic potential \cite{apa87} one has
$\Phi(x)=\left(1+\frac{1}{2}x\right)\mathrm{e}^{-x}$ and
\begin{equation}
\widetilde{V}(q) =
-4\pi Z_{1}e^{2} \,
\frac{q^{2}+2\alpha^{2}}{\left(q^{2}+\alpha^{2}\right)^{2}}.
\label{eq:29}
\end{equation}
In this case $\gamma=2$ \cite{ech89,lif98,ner13},
\begin{eqnarray}
G(s,u)
& = &
\frac{1}{16us} \,
\Bigg\{
\frac{1}{4}(25\eta\xi+13) \ln \frac{\eta^{2}+1}{\xi^{2}+1}
- \frac{25}{2}\eta\xi \, \ln\frac{\eta}{|\xi|}
\nonumber
\\
& &
\mbox{}
+ \frac{1}{3}(9\eta\xi+7)
\left( \frac{1}{\xi^{2}+1} - \frac{1}{\eta^{2}+1} \right)
\nonumber
\\
& &
\mbox{}
+ \frac{1}{3}(\eta\xi+1) \left[ \frac{1}{\left(
\xi^{2}+1\right)^{2}}-\frac{1}{\left(\eta^{2}+1\right)^{2}}\right]
\Bigg\}
\label{eq:30}
\end{eqnarray}
and $C_{\infty}=13/8$ \cite{ner13}. The auxiliary variables $\xi$ and
$\eta$ have been defined above.


\section{Results and discussion}
\label{sec:4}

Using the theoretical findings of Sections~\ref{sec:2} and \ref{sec:3}
we present here the numerical results for the Yukawa and hydrogenic
potentials. Protons ($Z_{1}=+1$) and antiprotons ($Z_{1}=-1$) are
considered along with a wide range of ion velocities, $v$, and a fixed
one-electron radius $r_{\mathrm{s}}=1.6$ ($v_{\mathrm{F}}\simeq
1.2v_{0}$, where $v_{0}$ is the Bohr velocity). Exact screening
parameters have also been computed for the same combinations of
$r_{\mathrm{s}}$, $Z_{1}$ and $v$. To this end, phase shifts were
evaluated by solving numerically the radial Schr\"{o}dinger equation for
the Yukawa and hydrogenic potentials, and inserting the phase shifts
into Eq.~(\ref{eq:10}). Then, a self-consistent iterative procedure
adjusted the value of $\alpha(v)$ so that the ensuing $\delta_{\ell}$
satisfy the exact dynamic FSR, Eqs.~(\ref{eq:9}) and (\ref{eq:10}).

\begin{figure*}[htbp]
\begin{center}
\includegraphics[width=82mm]{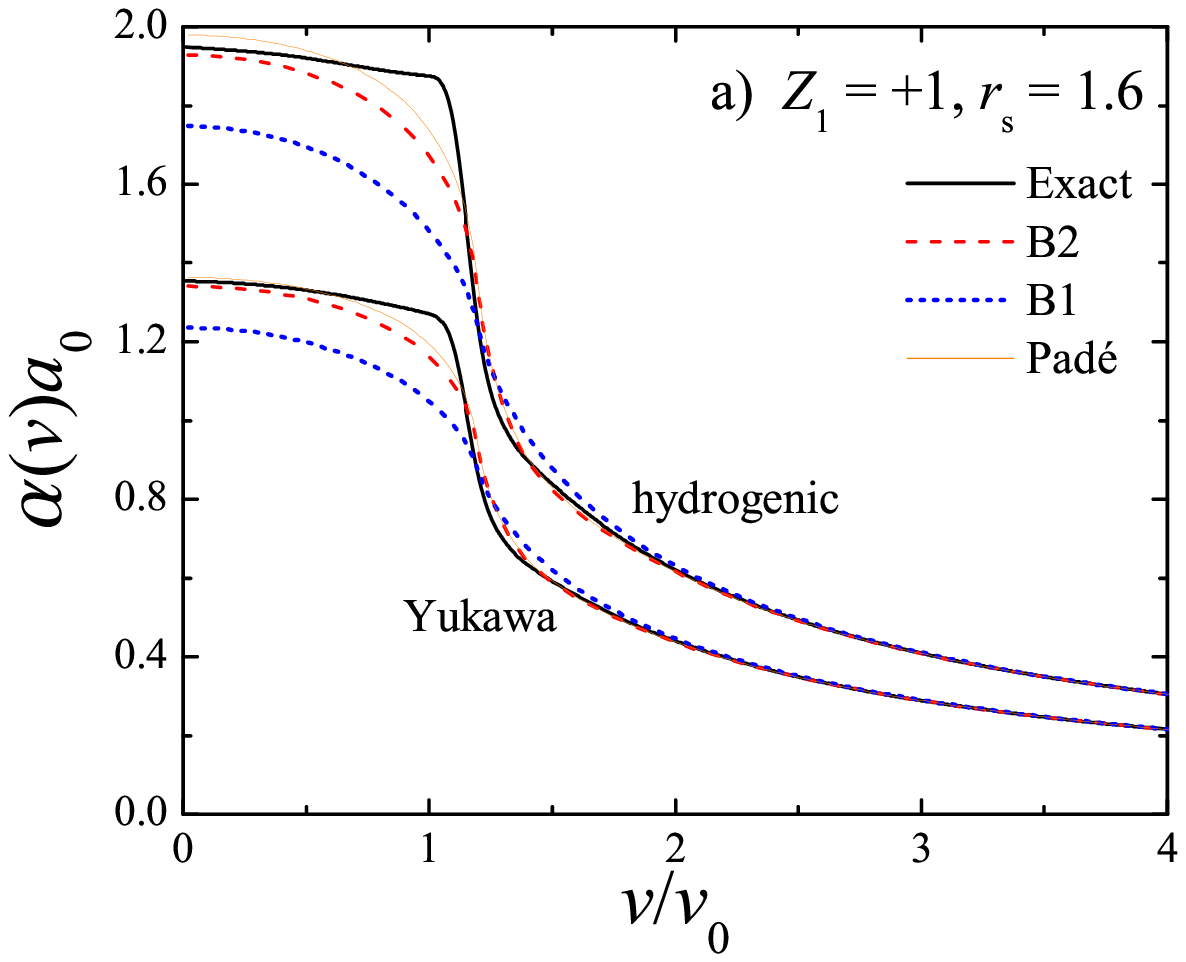}
\hspace*{2mm}
\includegraphics[width=82mm]{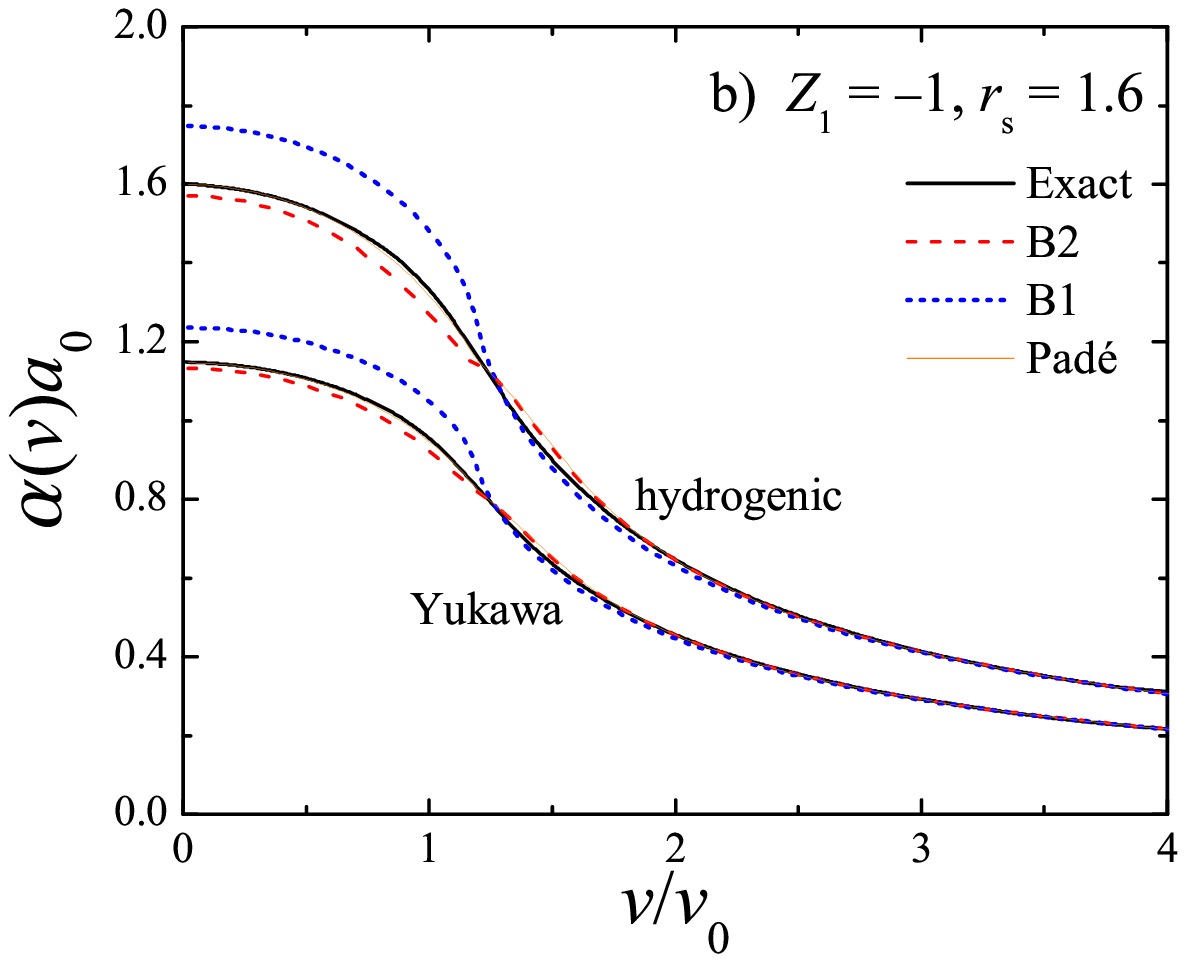}
\end{center}
\caption{Screening parameter $\alpha(v)a_{0}$ as a function of the ion
velocity $v/v_{0}$ for $r_{\mathrm{s}}=1.6$,
a) $Z_{1}=+1$, b) $Z_{1}=-1$,
calculated with the Yukawa and hydrogenic potentials.
The dotted curves corresponding to $\alpha_{\mathrm{B}1}(v)a_{0}$,
Eq.~(\ref{eq:24}).
The dashed curves indicate the solutions of
Eq.~(\ref{eq:20}), i.e.\ the B2 approximation.
The solid curves are the exact screening parameters calculated with
Eqs.~(\protect{\ref{eq:9}}) and (\protect{\ref{eq:10}}).
Eq.~(\protect{\ref{eq:pade}}) [i.e., the Pad\'{e} approximant of
Eq.~(\protect{\ref{eq:20}})] is plotted as thin solid curves.
}
\label{fig:1}
\end{figure*}

Fig.~\ref{fig:1} displays the dynamic screening parameter $\alpha(v)$
pertaining to the studied interaction potentials as a function of the
ion velocity $v$. Shown are the predictions of the B1 (dotted curves)
and B2 (dashed curves) approximations, given by Eqs.~(\ref{eq:24}) and
(\ref{eq:20}), respectively, and the exact screening parameters (solid
curves). It should be emphasized that, unlike the B1 screening parameter
$\alpha_{\mathrm{B1}}(v)$, the B2 approximation introduces a dependence
of $\alpha (v)$ on $Z_{1}$ and correctly predicts that at small
velocities $\alpha(v)>\alpha_{\mathrm{B1}}(v)$ if $Z_{1}>0$ and
$\alpha(v)<\alpha_{\mathrm{B1}}(v)$ if $Z_{1}<0$. At high velocities
($v\gtrsim 2v_{0}$) the perturbative and nonperturbative screening
parameters are in excellent agreement and this regime is accurately
approximated by the asymptotic Eq.~(\ref{eq:26}). In the opposite and
most unfavorable situation of intermediate and low velocities,
$v\lesssim 2v_{0}$, the B2 approximation deviates from the
self-consistent results of the exact dynamic FSR but it improves
significantly upon the $Z_{1}$-independent B1 approximation,
$\alpha_{\mathrm{B1}}(v)$. Furthermore, the B2 approximation is more
accurate for antiprotons (Fig.~\ref{fig:1}b) and the respective
screening parameters agree quite well with the exact treatment even at
low velocities. Similar trends have been reported in \cite{ner13} for
static ions immersed in a DEG.

The exact dynamic screening parameter of $Z_{1}=+1$ (Fig.~\ref{fig:1}a)
displays a conspicuous large negative slope close to the Fermi velocity
($v\simeq v_{\mathrm{F}}\simeq 1.2v_{0}$). In fact the behaviour of
$\partial\alpha(v)/\partial v$ is connected, through Eqs.~(\ref{eq:9})
and (\ref{eq:10}), to the derivative of the dynamic phase shifts,
$\partial\Delta_{\ell}(v)/\partial v$. When $v\to v_{\mathrm{F}}$ the
latter quantity is finite if $\ell\geqslant 1$ but exhibits a
logarithmic singularity if $\ell=0$, $\partial\Delta_{0}(v)/\partial v
\sim \lambda_{\mathrm{sc}}\ln |v-v_{\mathrm{F}}|$. Here
$\lambda_{\mathrm{sc}}=[\delta_{0}(q)/q]_{q\to 0}$ is the scattering
length \cite{lan81}. The exact theory of low-energy elastic scattering
(see, e.g., \cite{lan81}) predicts a finite scattering length for
protons and antiprotons, hence $\partial\alpha(v)/\partial v$ is
singular for both types of projectiles. However, $\lambda_{\mathrm{sc}}$
is about two orders of magnitude larger for protons than for antiprotons
and the singular behaviour of $\partial\alpha(v)/\partial v$ is clearly
visible only for $Z_{1}=+1$. The physical origin of this singularity
should be traced to the interaction of the positive ion (with
$v=v_{\mathrm{F}}$) with the electrons close to the Fermi surface in
which case the relative velocity of the particles can be small. As shown
in \cite{ner05} the cross section of the electron-ion interaction
strongly increases for these low-energy (in the relative frame of
reference) scattering events with $\ell=0$. This is the so-called
resonance scattering investigated first by Wigner and by Bethe and
Peierls (see, e.g., \cite{lan81}).

Finally, we have examined the Pad\'{e} approximant of order [1/1] to the
second-order Born series studied above. Applying this approximant one
finds, instead of Eq.~(\ref{eq:20}),
\begin{equation}
\alpha(v) =
\alpha_{\mathrm{RPA}} \big[ \gamma F(s) \big]^{1/2}
\left[
1 - \frac{\pi\chi^{2}}{2\gamma} Z_{1} \frac{G(s,u)}{F(s)}
\right]^{-1/2}.
\label{eq:pade}
\end{equation}
The resulting dynamic screening parameters are depicted in
Fig.~\ref{fig:1}. Comparing the different curves displayed in
Fig.~\ref{fig:1} one concludes that the [1/1] Pad\'{e} approximant
improves the agreement between the B2 approximation and the exact
results. The improvement is remarkable for antiprotons where the
respective screening parameters agree excellently with the
self-consistent treatment based on the dynamic FSR in the whole velocity
range.

In conclusion, we have proposed a method to calculate the dynamic
screening parameter for an ion moving in a DEG based on the B2
approximation for the FSR. The developed approach furnishes a simple and
computationally inexpensive scheme to incorporate the effects of the
non-linear ion-solid coupling in the quantum formulation of scattering
processes, which is an important prerequisite to describe accurately the
non-linear screening and energy loss of ions in solids.


\section*{Acknowledgements}

The work of H.B.~Nersisyan has been supported by the State Committee of
Science of the Armenian Ministry of Higher Education and Science
(project no.~13-1C200).
J.M.~Fern\'{a}ndez-Varea thanks the financial support from the
Generalitat de Cata\-lunya (project no.\ 2009~SGR~276).



\end{document}